\documentclass[aps,prb,reprint,noeprint,superscriptaddress]{revtex4-2}
\usepackage{graphicx}
\usepackage{hyperref}
\usepackage{natbib}

\usepackage{bm}
\usepackage{color}

\begin{document}


\title{{Suppression} of transport anisotropy at the Anderson localization transition in three-dimensional anisotropic media}


\author{Antton Goïcoechea}
\affiliation{Department of Physics and Astronomy, University of Manitoba, Winnipeg, Manitoba R3T 2N2, Canada}

\author{Sergey E. Skipetrov}
\affiliation{Univ. Grenoble Alpes, CNRS, LPMMC, 38000 Grenoble, France}

\author{John H. Page}
\email[]{john.page@umanitoba.ca}
\affiliation{Department of Physics and Astronomy, University of Manitoba, Winnipeg, Manitoba R3T 2N2, Canada}


\date{\today}

\begin{abstract}
We study {the} transport of classical waves through three-dimensional (3D) anisotropic media close to the Anderson localization transition.  Time-, frequency-, and position-resolved ultrasonic measurements {are} performed on anisotropic slab-shaped mesoglass samples to probe the dynamics and {the} anisotropy of the multiple scattering halo, and hence to investigate the influence of disorder on the nature of wave transport and its anisotropy.  These experiments allow us to address conflicting theoretical predictions that have been made about whether or not the transport anisotropy is affected by the interference effects that lead to Anderson localization.  We find that the transport anisotropy is significantly reduced as the mobility edge is approached{---a behavior similar} to {the one} predicted recently for matter waves in infinite anisotropic 3D media.
\end{abstract}


\maketitle


The transport of a wave{---}quantum or classical{---}through a disordered medium can be inhibited by interference effects, a phenomenon known as Anderson localization \cite{Anderson1958,abrahams201050,sheng2006introduction}. In three-dimensional (3D) systems, and only in 3D for the orthogonal symmetry class {for which no particular symmetry is broken}, a transition occurs between the extended regime (diffusive {transport}) and the localization regime {(exponentially suppressed transport)} when the disorder is increased \cite{Abrahams1979}. {For classical waves, the} localization regime is difficult to reach due to the very strong scattering needed to observe it; typically the scattering strength $k \ell_s$ has to be close to 1 (where $k$ is the wave {number} and $\ell_s$ is the scattering mean free path) according to the Ioffe-Regel criterion {\cite{ir1960,skip18}}. Anderson localization might be easier to achieve in strongly anisotropic {media} because {of the reduced} effective {dimensionality} \cite{Abrikosov1994,Kaas2008}. In such {media}, the scaling {behavior is expected to} be the same as in isotropic systems, and a single critical point remains. The effect of the anisotropy on the transition is nevertheless not trivial and there has been a lot of effort to incorporate it in numerical {models}, in particular in the context of cold atoms \cite{Pasek2017}. Additionally, early theoretical work using the self-consistent (SC) theory of localization predicted that the transport anisotropy (defined as the ratio of the diffusion tensor's components) is not affected by interference effects \cite{Wolfle1984,Bhatt1985}. This view has been challenged in Ref.~\cite{Piraud2014}, in which the authors use a more sophisticated version of the SC theory (see also Ref.~\cite{Yedjour2010}) and predict that the transport anisotropy {is significantly reduced if not} vanishingly small at the mobility edge. The {question concerning the} influence of the anisotropy on the transport of waves in very strongly scattering disordered systems is therefore not {settled}, motivating us to investigate {it} experimentally using ultrasound.

Ultrasonic waves have proven to be advantageous to observe localization \cite{Hu2008,Faez2009,Hildebrand2014,Aubry2014,Cobus2016,Cobus2018}, whereas 3D localization of light remains elusive \cite{Skipetrov2014,Skipetrov2016Page}. Because {we work with excitations of energy much exceeding both the single phonon energy $h f$ and the thermal energy $k_B T$ at room temperature $T$, our experiments can be interpreted using a classical wave equation and do not require low $T$ to be meaningful. This is an important advantage with respect to} quantum waves {(electrons or cold atoms) and allows us to relax} all the constraints inherent to {low-temperature} experiments. We also do not have to worry about interactions, present for electrons {and cold atoms and having} drastic effects on the transport properties (many-body localization, for example), {provided that we stay in the linear regime of wave propagation}. The length scales involved are also arguably more convenient since for typical ultrasonic experiments the wavelength is of the order of a {millimeter for the wave frequency $f \sim 1$ MHz}. The characteristic lengths in the samples are therefore between a fraction of a millimetre and a few centimetres, whereas the length scales in {optical} or matter wave experiments are orders of magnitude {shorter}. The amplitude and phase of acoustic signals can also be measured directly, which is generally very difficult in optics (in most experiments only the intensity is measured). Finally, time-, frequency-, and position-resolved ultrasonic experiments have been designed to measure transport quantities without the measurements being hindered by absorption \cite{Page1995,Cobus2018}. Using these techniques, we study experimentally{, close to the Anderson localization transition, two anisotropic samples with differently oriented structural anisotropy,} and focus in particular on the evolution of the transport anisotropy {upon approaching} the mobility edge. Our experiments provide a clear answer to the question of whether or not the transport anisotropy is significantly modified as the Anderson transition is approached.

{Our slab-shaped} samples {are} made of elongated, roughly ellipsoidal, aluminium particles brazed together. The method to {fabricate} the samples is covered in detail in Refs.~\cite{hu2006localization,cobus2016anderson,kerherve2018transport}. The links between the particles strongly influence the scattering strength of the {resulting} network. Once {a} sample is {fabricated}, it is possible to {enhance sound scattering in it} by etching {the} links by putting the sample in hydrochloric acid. {Etching makes} it possible to reach the critical regime {near} the Anderson localization transition. By comparing the highly reflective surface area of the polished outside surfaces with the highly reflective surface area of the inside cut surfaces of test samples, using the image-analysis software {{\tt ImageJ} \cite{imagej}} to measure the fraction of the total image area due to aluminium, we verified that the etching process is uniform.

The first sample investigated (A1) {is} made {of} small aluminium particles (approximately {3 mm $\times$ 1.5 mm $\times$ 1.5 mm}) aligned along the {$z$} axis {that is perpendicular to the slab}. After etching, the density of this sample is 1.269 g{/cm$^{3}$} (the aluminium volume fraction is approximately $47 \%$). The second sample (A2) consists of bigger aluminium particles (approximately {4 mm $\times$ 2 mm $\times$ 2 mm}) with the long axis of the particles aligned along {the $y$ axis} in the plane of the slab. The density of this sample {after etching} is 1.002 g{/cm$^{3}$} (the aluminium volume fraction is approximately $37 \%$). The scattering strength $k \ell_s$ in strongly scattering acoustic samples can be estimated from measurements of the average {transmitted} wave field \cite{Page1996,Cowan1998}. In our anisotropic samples, the scattering is so strong that the {challenge} of separating the average field from the multiply scattered waves makes the estimation difficult. Nonetheless, the data suggest that $k \ell_s \sim 1$ for A1e (where `e' stands for `etched'), and $k \ell_s \sim 2$ for A2e at the frequencies of interest. These values indicate very strong scattering and are close to the Ioffe-Regel criterion of localization $k \ell_s \sim 1$ {\cite{ir1960,skip18}}.

To study the transport of ultrasound in our anisotropic samples, we measure the {time dependence of the transmitted intensity profile for a point-like excitation at the origin $x = y = z = 0$}. This quantity has been measured previously in isotropic samples made of aluminium beads to investigate Anderson localization of ultrasonic waves without being obscured by absorption \cite{Hu2008,Cobus2018}. Experimentally, the {transmitted acoustic} field {$\psi(\bm{\rho},t)$} is measured at 9 detector positions near the sample surface {$z = L$:} ${\bm{\rho} = } (x,y)=\{(0,0),\allowbreak (\pm 15,0),\allowbreak (\pm 20,0),\allowbreak (0,\pm 15),\allowbreak (0,\pm 20)\}$ mm, for {$55 \times 55 = 3025$ different} source positions {at the opposite slab surface $z = 0$. For each measurement, the origin of the reference frame is fixed at the source position.} The separation between the source positions on the grid is equal to about one wavelength, enabling averaging over source positions to be done to estimate ensemble averages. The average transmitted intensity {$I(\bm{\rho},t) \propto \langle |\psi(\bm{\rho},t)|^2 \rangle$} is then used to calculate the transverse width squared:
\begin{equation}
w^2_{\bm{\rho}} (t) = -\rho^2/\ln \left[ \frac{I(\bm{\rho},t)}{I(0,t)} \right],
\label{eq:width squared}
\end{equation}
where $t$ denotes propagation time through the sample and $\rho=\sqrt{x^2+y^2}$. In the diffusive regime {and} for isotropic transport, {$I(\bm{\rho},t) = I(\rho,t)$ and} the width squared is simply $w^2_{\rho} (t) = 4 D t$, {where $D$ is the diffusion coefficient of ultrasound}. In the localization regime, the wave is confined in the transverse direction and the width squared saturates at long times \cite{Cherroret2010}. The width squared is therefore a direct measure of the dynamic spread of the wave through the sample and {of} its confinement by localization effects, if present.

We are mainly interested in the variation of the transport {anisotropy} as the mobility edge is approached. In the diffusive regime, {the anisotropy can be quantified by} the ratio of the different {elements $D_{xx}$, $D_{yy}$, $D_{zz}$} of the (diagonal) diffusion tensor $D$, irrespective of whether the medium is finite or not. As one gets closer to the {localization} transition, this {way of quantifying transport anisotropy} becomes ill-adapted to experimental studies, since the ratio of {the} elements of $D$ is not directly accessible and may not even be uniquely defined in finite-sized samples, where $D$ becomes position dependent. {Moreover,} for sample A1, the transport anisotropy {is not directly visible in the intensity profile $I(\bm{\rho},t)$ and requires a careful comparison with theoretical models to be assessed. In contrast, for} sample A2, the anisotropy is in the plane of the slab {and is thus expected to manifest itself by a deviation of $I(\bm{\rho},t)$ from a circularly symmetric shape, leading to a difference between $w^2_{x}(t)$ and $w^2_{y}(t)${,} which are short-hand notations for $w^2_{\bm{\rho}}$ with $\bm{\rho} = (x,0)$ and $\bm{\rho} = (0,y)$, respectively.} We can therefore define {a parameter $\mathcal{A}_{\rho}$ characterizing the transport anisotropy as
\begin{equation}
\mathcal{A}_{\rho} = \left. \lim\limits_{t \to \infty} \frac{w^2_{y} (t)}{w^2_{x} (t)} \right|_{x = y = \rho}.
\label{eq:a}
\end{equation}
In the diffusive regime this is equivalent to {$\mathcal{A}_{\rho} = D_{yy}/D_{xx}$}}.

\begin{figure}
\includegraphics[width=0.48\textwidth]{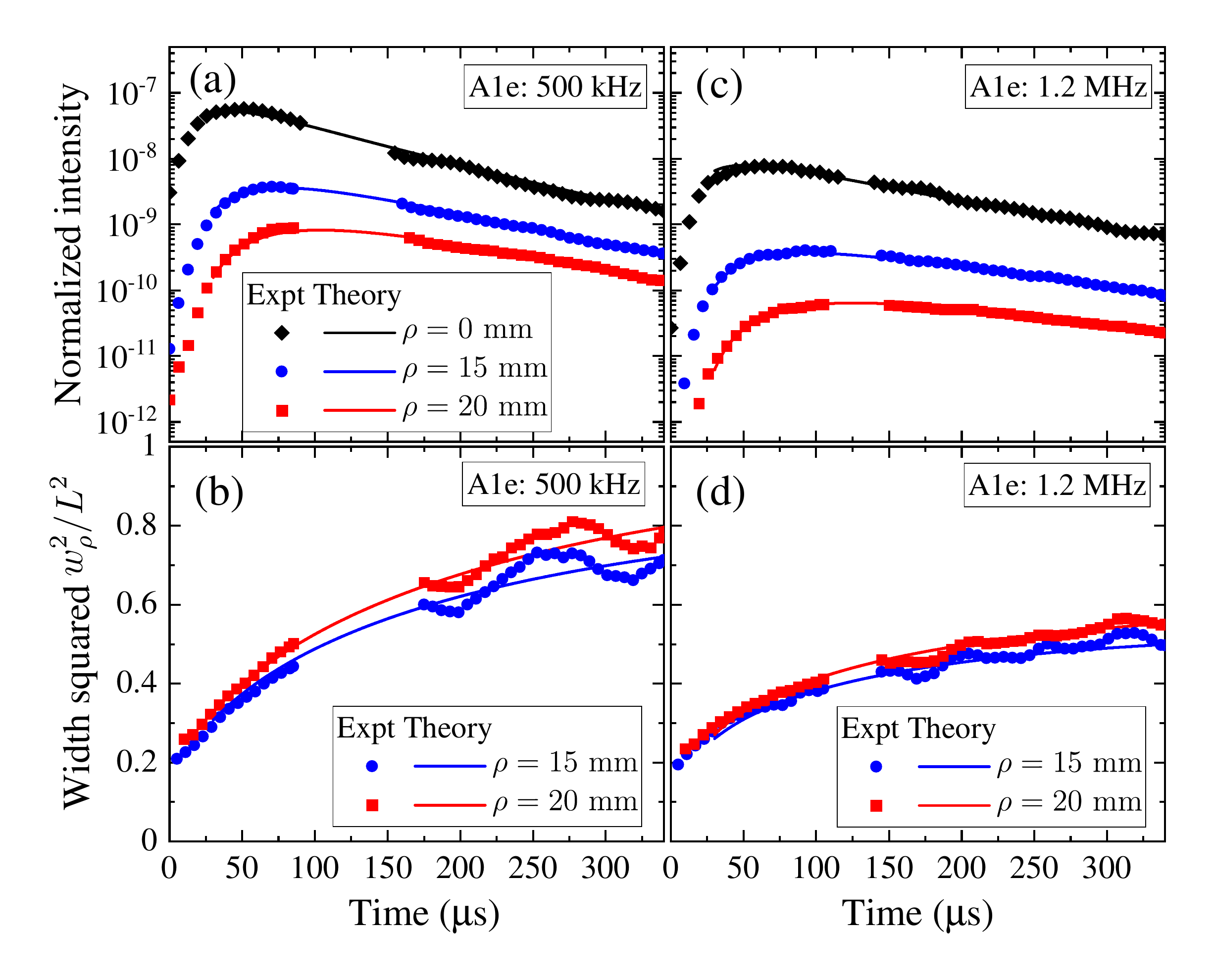}
\caption{\label{TOF w2 A1e}(a), (c) Time of flight (TOF) profiles $I(\rho,t)$ and (b), (d) width squared $w^2_{\rho} (t)$ as {functions} of time for the etched sample A1e at 0.5 MHz and 1.2 MHz, respectively (bandwidth of 50 kHz). {Symbols} represent experimental data and {solid lines} represent {fits with the isotropic} SC theory. The legend for (a) also applies to (c). The missing data near 125 $\mu$s are not shown because of contamination by a spurious electronic pickup pulse. At 0.5 MHz, the {best fits are obtained} for $L/\zeta = 0.5$ [$k \ell_s > (k \ell_s)_c $]; at 1.2 MHz, we obtain $L/\xi=2.5$ [$k \ell_s < (k \ell_s)_c $]. The good {quality of fits suggests} that the transport is {diffusive (although strongly renormalized by localization effects)} at 0.5 MHz and that {Anderson} localization is {reached at} 1.2 MHz. {The good fit quality also suggests that the transport anisotropy is greatly reduced at both frequencies.}}
\end{figure}

We first present the results for sample A1, for which the long axis of the particles is {parallel to the $z$ axis and $D_{xx} = D_{yy} = D_{\perp} \ne D_{zz}$}. Before etching, the transport near {$f = 1$ MHz is} diffusive ($w_{{\rho}}^2 \propto t$), with significant anisotropy. {Fits of} the experimental data for $I(\rho,t)$ and $w_{{\rho}}^2(t)$ with the {diffusion theory yield} the transport anisotropy {$D_{zz}/D_{\perp} \simeq 2.5$}. {In} contrast, after etching, {a} completely different {behavior is} found, with {$w_{\rho}^2(t)$} bending over significantly {with time} at all {investigated frequencies $f$} from 0.4 to 1.3 MHz. Representative data at 0.5 and 1.2 MHz are shown in Fig.~\ref{TOF w2 A1e}, where {at both frequencies $w_{\rho}^2(t)$ tends to} saturate with time at a low value $w^2_{{\rho}}(t \rightarrow \infty )/L^2 \leq 1$. Note that {SC theory predicts $w^2_{\rho}(t \rightarrow \infty )/L^2 \sim 1$ for thick isotropic samples at the mobility edge} \cite{Cherroret2010}. Thus, Fig.~\ref{TOF w2 A1e} {clearly indicates} that the {transport} is not diffusive for sample A1e, and that localization is close to taking place in this frequency range, with the tendency of the width squared to saturate being more pronounced at 1.2 MHz. At both frequencies, $w^2_{\rho} (t)$ depends on the value of $\rho$, which is the expected {behavior near a} mobility edge \cite{Hu2008}.

The absence of a transport theory for anisotropic media {of finite size} makes it challenging to {determine with certainty} whether {Anderson localization} has been reached {in sample A1e}. Nevertheless, the significant deviations from classical diffusion observed in Fig.\ \ref{TOF w2 A1e} are likely to be accompanied by a suppression of transport anisotropy according to the recent theoretical work \cite{Piraud2014}. It is then tempting to {fit} the data with the {SC theory for isotropic media}. Indeed, if the anisotropy does disappear, then we could expect the theory to work. Note that for isotropic systems, comparison of this theory with experimental data for $I(\rho,t)$ and $w^2_{\rho} (t)$ has enabled the localized and extended regimes to be distinguished, and mobility edges identified \cite{Hu2008,Cobus2016,Cobus2018}. Proximity to a mobility edge is conveniently assessed by the localization length $\xi$ when $k \ell_s < (k \ell_s)_c$ (localization regime) and the {correlation length of fluctuations} $\zeta$ when $k \ell_s > (k \ell_s)_c $ (diffusive regime) \cite{Cobus2018}. {Here $(k \ell_s)_c \sim 1$ is the value of $k \ell_s$ at the mobility edge} \footnote{See, for example, Ref. \cite{Cobus2018} for the relationships between $\xi$, $\zeta$, $k \ell_s$ and $(k \ell_s)_c $.}. Both $\xi$ and $\zeta$ diverge at the mobility edge.

{Fits of excellent quality are obtained} when {the} isotropic SC theory is {used to fit} the experimental data for sample A1e, {see Fig.~\ref{TOF w2 A1e}}. At 0.5 MHz [Fig.~\ref{TOF w2 A1e}(a) and (b)], the best {fits are} found for $L/\zeta = 0.5$ [$k \ell_s / (k \ell_s)_c = 1.003 $], consistent with the system being still in the extended regime but very close to the mobility edge at this frequency. We can also infer, since the fits are good, that the transport anisotropy is greatly reduced this close to the mobility edge {(recall that  $D_{zz}/D_{\perp} \simeq 2.5$} for this sample before etching). {At 1.2 MHz [Fig.~\ref{TOF w2 A1e}(c) and (d)]} the best {fits are}  found for $L/\xi=2.5$ [$k \ell_s /(k \ell_s)_c $ = 0.988], indicating {Anderson localization of ultrasound}. Thus, not only do the good fits obtained using the isotropic SC theory with $L/\xi=2.5$ indicate that {Anderson} localization has been achieved, but they also show that the transport anisotropy must be so greatly reduced in the localization regime that it has negligible impact on the fit quality.

\begin{figure}
\includegraphics[width=0.48\textwidth]{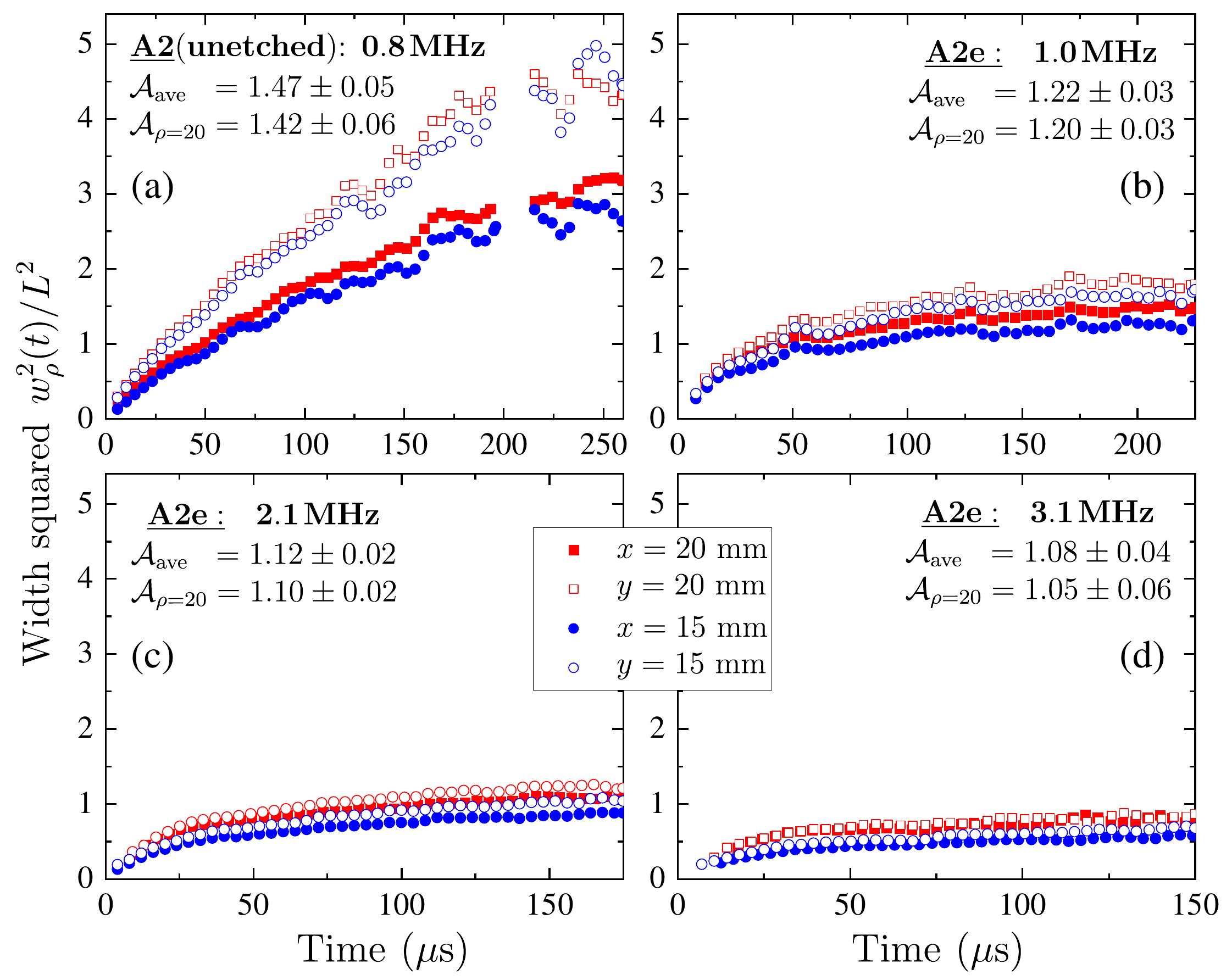}%
\caption{\label{w2 4freq exp}Comparison of width squared for samples A2 and A2{e} at several frequencies, up to the maximum frequency at which the transmitted signals were large enough for $w^2_{{\bm{\rho}}} (t)$ to be measured. As the {frequency} increases, the widths squared bend over more and more, and the difference between {$w^2_{x} (t)$ and  $w^2_{y} (t)$} decreases, giving a clear visualization of the reduction in transport anisotropy as the Anderson transition is approached.}
\end{figure}

To obtain {information} on the evolution of the transport anisotropy in the approach to localization {in a more direct way}, we now {turn to sample A2}, for which the anisotropy is in the plane of the slab (particles' major axis {parallel to the $y$ axis}). The {behavior} of the width {squared,} as a function of structural disorder and frequency, is illustrated in Fig.~\ref{w2 4freq exp}, which shows a comparison of the widths squared for both the unetched and etched samples, at four different frequencies and for two $\rho$ values. For the unetched sample A2 (panel (a)), the width squared at 800 kHz shows only modest deviations from the linear growth in time that is characteristic of purely diffuse {behavior}, and the separation between the width squared for {$\bm{\rho} = (x = \rho, 0)$ and $\bm{\rho} = (0, y = \rho)$} is large ({compare} closed and open symbols {of} the same {color} and type). For the etched sample A2e, {the} transverse spreading of wave energy is much slower, with {$w_{\bm{\rho}}^2(t)$} bending over significantly as time increases, at all frequencies. {In} particular, as the frequency is increased, $w_{x}^2(t)$ and $w_{y}^2(t)$ bend over more and more, and {get} closer and closer together. Notably, at 3.1 MHz, the width squared indicates substantial confinement of the wave energy and the values corresponding to the two different directions are very close to each other, showing that the transport anisotropy has almost vanished. These observations demonstrate that the etching has increased the disorder inside the sample, and that it is reasonable to expect that the mobility edge is close to 3.1 MHz (because the width squared looks like it is about to saturate at long times). Thus, as {the} width squared approaches {the behavior expected} at the mobility edge ({saturation} at a constant value $\sim L^2$ as $t \rightarrow \infty$), the anisotropy {is} progressively reduced and appears to have almost disappeared at 3.1 MHz.

\begin{figure}%
\includegraphics[width=0.42\textwidth]{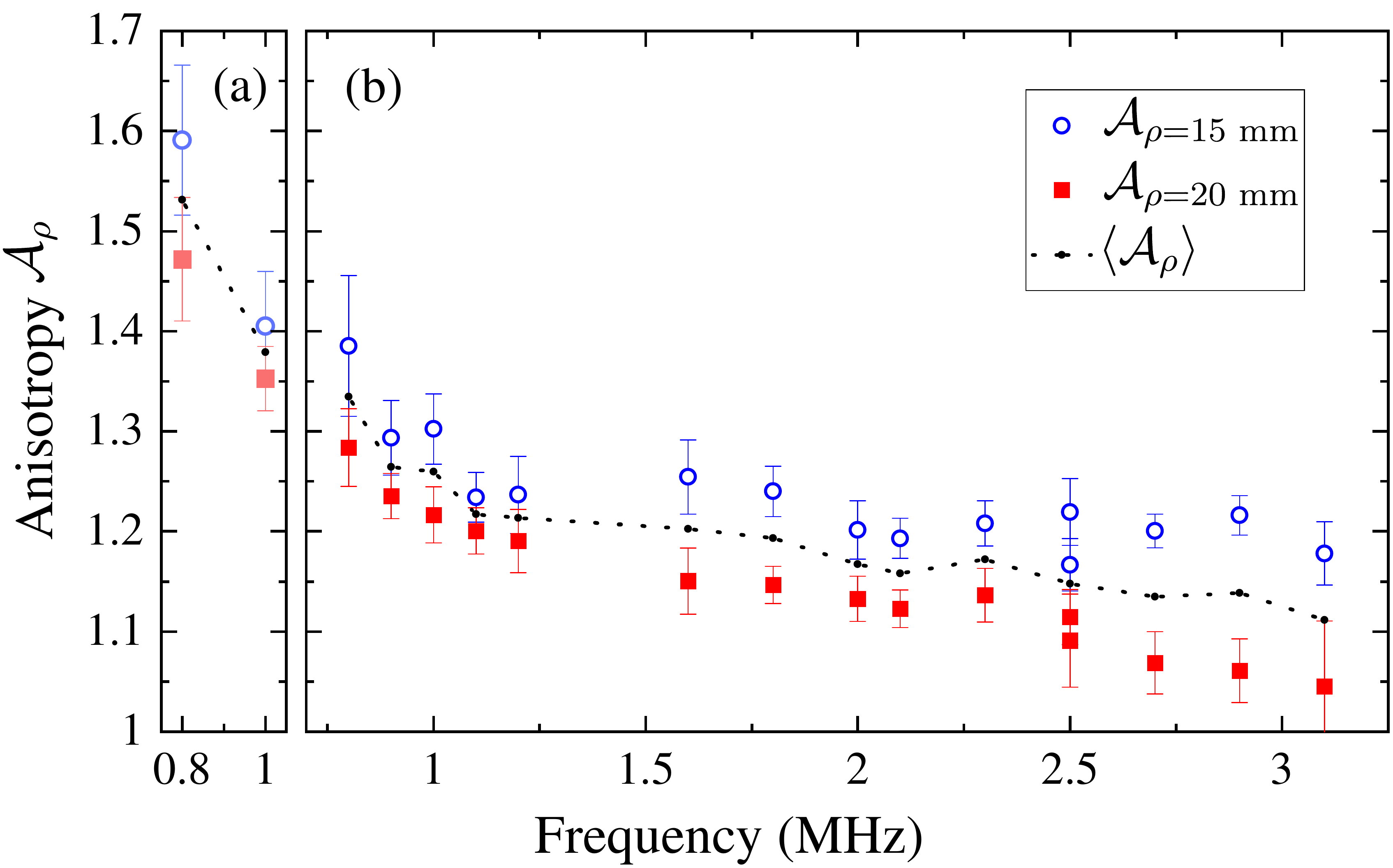}
\caption{Comparison of the transport anisotropy {defined by Eq.\ (\ref{eq:a})} as a function of frequency for samples A2 in panel (a) and A2e in panel (b). The anisotropy measured separately for $\rho = 20$ mm and $\rho = 15$ mm is presented with symbols while the average anisotropy for transverse distances of 15 and 20 mm is shown by the dotted lines. As the frequency increases, leading to an increase in the effective disorder (see the effect on the width squared in Fig.~\ref{w2 4freq exp}), the anisotropy decreases, with $\mathcal{A}_{{\rho}}$ approaching 1 (no anisotropy) at the highest frequency for which data could be obtained. The decrease in anisotropy is more pronounced at the greater transverse distance of $\rho = 20$ mm.}
\label{anisotropy A2&A2m}
\end{figure}

The anisotropy {$\mathcal{A}_{\rho}$} is shown in Fig.~\ref{anisotropy A2&A2m} as a function of frequency for both the unetched and etched samples. {The} evolution of the anisotropy {is indicated by symbols for two transverse distances $\rho =$ 15 and 20 mm}, while the dotted curve denotes their average. This direct measurement {clearly} demonstrates that there is a substantial decrease of the anisotropy as the frequency, and hence the effective disorder strength, is increased. In particular, it is interesting to look at the differences in the measured anisotropy with transverse distance. At $\rho = 20$ mm, the anisotropy continues to decrease as the {highest} attainable frequency of 3.1 MHz is reached and appears to extrapolate to 1 at a slightly higher frequency. (At 3.1 MHz, $\mathcal{A}_{{\rho=20}} = 1.045 \pm 0.065$, and is actually consistent with 1 even at this frequency.) However, at $\rho = 15$ mm, the anisotropy is larger at all frequencies, and {its} decrease in the upper part of the accessible frequency range is less obvious. These data strongly suggest that as the localization transition is approached, the {shape} of the multiply scattering halo {$I(\bm{\rho},t)$} changes: it remains somewhat anisotropic near ${\bm{\rho}} = 0$ but its anisotropy appears to nearly vanish at large {distances $\rho$}. It is tempting to consider that large {distances $\rho$} may capture more closely in a finite sample the situation encountered in {the} infinite medium, where the SC theory has predicted that the transport anisotropy {is strongly suppressed or even vanishes} at the mobility edge \cite{Piraud2014}. Thus, {our experiments} seem to confirm the predictions of Piraud \textit{et al.} \cite{Piraud2014}. In addition, our results {demonstrate} that in real samples, which are necessarily {of finite size}, the way in which {the suppression} of the anisotropy shows up is more complex.

In summary, we have shown that samples made of elongated aluminium particles, aligned along a given direction, are suitable to investigate Anderson localization of ultrasound in anisotropic media. To assess the transport anisotropy, we have proposed a generalized definition in terms of the anisotropy in the width of the multiple scattering halo, which reduces to the standard definition as a diffusion coefficient ratio in diffuse media but has the advantage of being directly measurable experimentally for finite samples in both the {extended} and {localization} regimes. Thus, we have demonstrated that the transport anisotropy is significantly reduced as the mobility edge is approached {and that the wave transport tends} towards isotropic behavior very close to the localization transition. This work constitutes the first experimental observation of this {behavior}, and supports the {more recent} SC theory prediction for infinite media \cite{Piraud2014}. Our findings indicate that {the} ``ellipsoidal'' cutoff in the SC theory {used in Refs.}~\cite{Wolfle1984,Bhatt1985} may not be valid, as {the predictions it produces} are not consistent with our data. It is worth mentioning that our anisotropic samples {feature} a {much more gradual} approach to the localization regime as the frequency is varied {than what was observed in} previously studied samples made of aluminium beads \cite{Cobus2018}. This is likely due to the irregular shape and the polydispersity of the anisotropic particles, as well as to the control over the scattering strength (via the brazing conditions and post-fabrication etching of the samples) that enabled the critical regime to be {reached.} This result motivates further experimental work with such samples to explore {their behavior} near the mobility edge; assuming such experiments are successful and sufficient configurational averaging can be achieved, a finite-time scaling analysis could be performed in order to extract the critical exponent of the Anderson transition \cite{Lemarie2009,Beltukov2017}.

\begin{acknowledgements}
We are grateful for support from the Natural Sciences and Engineering Research Council of Canada’s Discovery Grant Program (RGPIN-2016-06042).
\end{acknowledgements}

\nocite{*}

\end{document}